\documentstyle [12pt]{article}

\begin{document}
\baselineskip 7.5 mm

\def\thefootnote{\fnsymbol{footnote}}
\baselineskip 7.5 mm

\begin{flushright}
\begin{tabular}{l}
UPR-656-T \\
hep-ph/9504418 \\
April, 1995
\end{tabular}
\end{flushright}

\vspace{20mm}

\begin{center}

{\Large \bf Improved Action Method for Analyzing Tunneling }
\\
\vspace{2mm}
{\Large \bf in Quantum Field Theory}
\\

\vspace{20mm}

\setcounter{footnote}{0}

Alexander Kusenko\footnote{ email address: sasha@langacker.hep.upenn.edu}
\\
Department of Physics and Astronomy \\
University of Pennsylvania \\
Philadelphia, PA 19104-6396 \\

\vspace{30mm}

{\bf Abstract}
\end{center}

We describe a new method which allows one to evaluate the
false vacuum decay rate for a general potential which may depend on an
arbitrary number of scalar fields.

\vfill

\pagestyle{empty}

\pagebreak

\pagestyle{plain}
\pagenumbering{arabic}
\renewcommand{\thefootnote}{\arabic{footnote}}
\setcounter{footnote}{0}

\pagestyle{plain}

\section{Introduction}

The problem of calculating the transition probability between two
inequivalent vacua in quantum field theory arises in different areas of
high-energy physics, cosmology, and condensed matter physics.
The semiclassical calculation of the false vacuum decay width was done in
Refs. \cite{vko,c,cc} for the case of a single scalar field, $\phi(x)$.  The
corresponding path integral is dominated by the field configuration
$\bar{\phi}(x)$ called the ``bounce'' and can be evaluated using the saddle
point method  \cite{c,cc}.
The bounce, being the stationary point of the Euclidean
action, is the non-trivial solution of the corresponding Euler-Lagrange
equation which obeys certain boundary conditions.

Then the transition probability per unit volume, in the semiclassical limit
\cite{cc} is

\begin{equation}
\Gamma/{\sf V}=\frac{1}{h^2} (S[\bar{\phi}])^2 e^{-S[\bar{\phi}]/\hbar}
\left |
\frac{\det'[-\partial_\mu^2+U''(\bar{\phi})]}{\det[-\partial_\mu^2+U''(0)]}
\right |^{-1/2}
\times (1+O(\hbar))
\label{one}
\end{equation}
where $S[\bar{\phi}]$ is the Euclidean action of the bounce, and $\det'$ stands
for the determinant with the zero eigenvalues omitted.

However, a number of practical applications require a generalization of
this method to the case that involves several scalar fields.
Although, from a theoretical point of view, such a generalization is rather
straightforward\footnote{If the potential has a continuum of degenerate
minima, then the determinant in equation (\ref{one}) acquires some
additional zero modes, so that the right-hand side of (\ref{one}) is, in
fact, divergent.   The false vaccum decay rate in this case was calculated
in Ref. \cite{ak}},  in practice it is impossible to find the bounce
except in trivial cases.   The problem is that the bounce is an
unstable solution of some system of non-linear differential equations which
cannot be solved analytically.  Small changes in the initial conditions
lead to significant changes in the shape of the solution. In the case of a
single
scalar field, $\phi(x)$, one can find the bounce numerically by imposing the
boundary condition on the value of the so called ``escape point'',
$\phi_e=\phi(0)$, and
constructing the ``overshoot'' and the  ``undershoot'' solutions which
envelope the bounce.   Due to the special topology of the one-dimensional
problem, one can then constrain the bounce to any given accuracy by varying
the ``escape'' point, the value of the field at the center of the bounce,
so that the ``overshoot'' and the ``undershoot'' solutions converge.
If, however, the scalar field has $n>1$ components, this approach
fails because an arbitrary path connecting the $\phi_e$ of the
``overshoot'' with that of the ``undershoot'' does not have to go through
the ``true'' escape point.

The bounce $\bar{\phi}(x)$ is defined as a stationary point of the action
$S[\phi]$ and is the solution of the variational equation
$\delta~S[\phi]=0$.  However, it is not a minimum of $S[\phi]$ but rather a
saddle point
\footnote{
The method proposed in Ref. \cite{chh} relies on the minimization of the
action on the lattice with respect to the variations of the bounce profile,
while maximizing the same action with respect to one particular variation,
namely the scale transformation.  Since the scaling corresponds to the
direction of the instability of the bounce \cite{derrick,c_comm},
one can argue that
it is desirable to minimize the action with respect to the variations
orthogonal to scaling, and at the same time maximize it with respect to
scaling.  However, in practice, it is nearly impossible to identify the
variations orthogonal to that related to scaling.  And, in general, an
iterative procedure of this kind does not converge to any limiting point.}.

If one could find some functional $\tilde{S}[\phi]$ for which the
bounce would be a minimum, then one could define the the function $\phi(x)$
on a lattice and minimize $\tilde{S}[\phi]$ with respect to random
variations of
$\phi(x)$.  This is the basic idea of our method.  We will show that by
adding some auxiliary terms to the action one can turn the saddle
point into a true minimum.  First, we derive some useful identities which
help identify the terms that may be added to the action.  And then we prove
that the saddle point of the original action is a minimum of the improved
action.

\section{General properties of the bounce}

Let us consider a quantum field theory with the scalar potential potential
$U(\phi_1, ... , \phi_n)$ which has a local minimum
at $\phi_1=\phi_2=...=\phi_n=0$, $U(0,...,0)=0$ as well as at least one
additional (local, or global) minimum at $\phi_i=\phi_i^e, \ i=1,2,...,n$;
$U(\phi_1^e,...,\phi_n^e)<0$.  In the semiclassical limit \cite{vko,c,cc},
the transition probability is proportional
to $\exp(-S[\bar{\phi}(x)])$, where $\bar{\phi}(x)=(\bar{\phi}_1(x), ...
,\bar{\phi}_n(x) )$
is the so called ``bounce'', a nontrivial
$O(4)$-symmetric \cite{c_comm} field configuration determined by the system
of equations

\begin{equation}
   \Delta \bar{\phi}_i(r)= \frac{\partial}{\partial \bar{\phi}_i}
U(\bar{\phi}_1, ... , \bar{\phi}_n)
\label{bounce_eq}
\end{equation}
with the following boundary conditions:

\begin{equation}
\left \{ \begin{array}{l}
    (d/dr) \bar{\phi}_i(r)|_{r=0}=0 \\  \\
   \bar{\phi}_i(\infty)=0
        \end{array} \right.
\label{boundary}
\end{equation}

As was explained in the introduction, the chaotic instability of the
solution of (\ref{bounce_eq}--\ref{boundary}) with respect to small
perturbations makes it impossible to solve the equation numerically.  We
therefore turn to the variational equation (for which (\ref{bounce_eq}) is
the consecutive Euler-Langrange equation):

\begin{equation}
\delta S=0
\label{dS}
\end{equation}
where $S$ is the Euclidean action for a scalar field $\phi(x)$:

\begin{eqnarray}
S & = & T+V \\  \nonumber \\
T & = & \int d^4 x \sum_{i=1}^4 \frac{1}{2}(\frac{\partial
\phi(x)}{\partial x_{i}})^2
\label{T}  \\ \nonumber \\
V & = & \int U(\phi(x)) d^4x
\label{V}
\end{eqnarray}

For any renormalizable theory\footnote{In the presence of some higher-order
(non-renormilizable) terms, equations (\ref{l2}) and (\ref{lT2}) are
modified accordingly to include the monomials of higher powers.  However,
the same method can be used to construct the improved action.}, the
potential term in the Euclidean action $V=\int U(x) d^4x$ can be written as

\begin{eqnarray}
V & = & V_2+V_3+V_4 \\
V_n[\phi] & = & \int U_n(\phi(x)) d^4 x
\end{eqnarray}
where  $U_n(\phi)$ is $n$-linear in $\phi$.  In other words,

\begin{equation}
V_n[\lambda \phi]  = \int U_n( \lambda \phi(x)) d^4 x =
\int \lambda^n U_n(\phi(x)) d^4 x  =  \lambda^n V_n[\phi]
\end{equation}
Here we used the fact that $\phi=0$ is the local minimum and therefore
no term, linear in $\phi$, is present in the potential.

By virtue of (\ref{dS}), the solution $\bar{\phi}(x)$ is the
stationary point of
the action.  Therefore, for any family of functions $\phi_\lambda(x)$ such
that $\phi_\lambda(x)|_{_{\lambda=1}}=\bar{\phi}(x)$, the corresponding action
$S[\phi_\lambda(x)]$ should have a zero derivative with respect to $\lambda$
at $\lambda=1$:

\begin{equation}
\frac{d}{d\lambda} S[\phi_{\lambda}(x)] \mid_{_{\lambda=1}} = 0
\label{dSl}
\end{equation}

In particular, one can choose

\begin{equation}
\phi^{(\lambda)}(x)=\lambda^p \bar{\phi}(\lambda^q x),
\label{scaling}
\end{equation}
where $p$ and $q$ are arbitrary real numbers.

Then

\begin{equation}
S[\phi^{(\lambda)}(x)]= \lambda^{2(p-q)} T+ \sum_{n=2}^4 \lambda^{pn-4q}V_n
\end{equation}

The equation (\ref{dSl}) implies

\begin{eqnarray}
2 [p-q] T+ \sum_{n=2}^4 [n p-4q] V_n & = & 0  \Rightarrow  \nonumber \\
\nonumber \\
p(2T+ 2V_2+3V_3+4V_4) -2 q (T+2V) & = & 0
\label{rel}
\end{eqnarray}

The relation (\ref{rel}) can be satisfied for all $p$ and $q$ if and only
if

\begin{eqnarray}
T+2V & = & 0
\label{TV} \\
2V_2+ V_3 & = & 0
\label{V2V3}
\end{eqnarray}
The first of these two identities (\ref{TV}) is well known and can be
derived \cite{derrick,c_comm} using the variation of the type
(\ref{scaling}) with $p=0, \ q=1$.

Another useful identity can be obtained by considering the following
variation about the bounce:

\begin{equation}
\left \{ \begin{array}{l}
\bar{\phi}_i(x) \rightarrow \phi^{(\lambda)}_i(x) = \bar{\phi}_i(x) -
\lambda \bar{\phi}_j(x)  \\
\bar{\phi}_j(x) \rightarrow \phi^{(\lambda)}_j(x) = \bar{\phi}_j(x) +
\lambda \bar{\phi}_i(x)  \\
\bar{\phi}_k(x) \rightarrow \phi^{(\lambda)}_k(x) = \bar{\phi}_k(x), \ \ k
\neq i,j \\
         \end{array}  \right.
\label{variation}
\end{equation}

It is convenient to write (\ref{variation}) as

\begin{equation}
z(x) \equiv \bar{\phi}_i(x)+ i \bar{\phi}_j(x) \rightarrow
z^{(\lambda)}(x)=e^{i
\lambda}(\bar{\phi}_i(x)+ i \bar{\phi}_j(x))
\label{z_var}
\end{equation}

Then it is clear that the kinetic term in the Euclidean action

\begin{eqnarray}
T=\int d^4 x \sum_{a,k} (\frac{d\phi_k(x_a)}{dx_a})^2 =  \nonumber \\
\\ \nonumber
\int d^4 x \sum_{a; \: k\neq i,j} (\frac{d\phi_k(x_a)}{dx_a})^2 +
\sum_a (\frac{d}{dx_a} \bar{z}) (\frac{d}{dx_a} z)
\label{kin_z}
\end{eqnarray}
is invariant under the transformation (\ref{z_var}), and therefore under
the transformation (\ref{variation}).

It is also clear that the transformation (\ref{variation}) preserves the
boundary conditions (\ref{boundary}).  In other words, if the constraints
(\ref{boundary}) are satisfied for $\lambda=0$, then they are also satisfied
for any non-zero value of $\lambda$.

We now apply the transformation (\ref{variation}) to the Euclidean action
and require that it have zero derivative with respect to the parameter
$\lambda$ at $\lambda=0$:

\begin{eqnarray}
0=(d/d\lambda) S[\phi_\lambda]_{\lambda=0}= (d/d\lambda)
(T+V)_{\lambda=0}= (d/d\lambda) V |_{\lambda=0}=  \nonumber \\
\nonumber \\
\int d^4x  ( \phi_j(x) \frac{\partial}{\partial \phi_i} -
\phi_i(x) \frac{\partial}{\partial \phi_j})U(\phi)|_{\phi(x)=\bar{\phi}(x)}
\end{eqnarray}

Thus for an arbitrary pair of $\phi_i$ and $\phi_j$ we obtain the
constraint:

\begin{equation}
\int d^4 x \det \left | \begin{array}{cc}
                    \frac{\partial}{\partial \phi_i} &
\frac{\partial}{\partial \phi_j}\\
                    \phi_i &  \phi_j
                   \end{array}  \right | U(\phi_1, ... ,\phi_n)
\mid_{_{\phi(x)=\bar{\phi}(x)}} = 0
\label{det_con}
\end{equation}

There are $(n-1)$ linearly-independent constraints of the form
(\ref{det_con}) for $(n-1)$ pairs of components of the $n$-component
field $\phi$.

\section{Improved Action method}

In the preceeding section we proved that the following quantities vanish
identically:

\begin{eqnarray}
\Lambda_1 & = & T[\bar{\phi}(x)]+2V[\bar{\phi}(x)] \label{l1} \\ & &
\nonumber \\
\Lambda_2 & = & 2 V_2[\bar{\phi}(x)] +V_3[\bar{\phi}(x)] \label{l2} \\ & &
\nonumber \\
\Lambda_3^{ij} &  = & \int d^4 x \  \det
           \left | \begin{array}{cc}
                    \frac{\partial}{\partial \phi_i} &
\frac{\partial}{\partial \phi_j}\\
                    \phi_i &  \phi_j
                   \end{array}  \right | U(\phi_1, ... ,\phi_n)
\mid_{_{\phi(x)=\bar{\phi}(x)}}  \label{l3}
\end{eqnarray}

In order to determine the bounce numerically,
we would like to find some functional $\tilde{S}[\phi]$ for which the bounce
$\bar{\phi}(x)$ would be a minimum.  Then one can discretize the problem by
defining the field $\phi(x)$ on a lattice.  The improved action,
$\tilde{S}[\phi]$ can then be minimized with respect to random perturbations of
the field $\phi$ until the limiting bounce is recovered to a required
accuracy\footnote{Clearly, it is not even necessary that
$\tilde{S}[\phi]$ be differentiable with respect to $\phi$ at
$\phi=\bar{\phi}(x)$.}.

We define $\tilde{S}[\phi]$ as the action $S[\phi]$ plus some auxiliary terms
which would vanish as $\phi(x)$ approaches $\bar{\phi}(x)$:

\begin{equation}
\tilde{S}[\phi]:= S[\phi]+ \sum_{n} (\alpha_{1,n} |\Lambda_1|^{p_n}+
\alpha_{2,n} |\Lambda_2|^{p_n}+\alpha_{3, ij , n} |\Lambda_3^{ij}|^{p_n}),
\label{imprvd}
\end{equation}
where $p_n, \ n=1,2,...$ are some positive numbers, and $\alpha_n>0$ are
some arbitrary Lagrange multipliers.

In particular, we will show that it suffices to take

\begin{equation}
\tilde{S}[\phi]:= S[\phi]+  \alpha_{1} |\Lambda_1|
\label{st}
\end{equation}
but in practice one can include other terms as well to speed up the
convergence of the iterative routine.

Now we want to prove the following statement:

\vspace{3mm}

\noindent {\bf Theorem 1.}
{\em If $\delta S [\bar{\phi}(x)]=0$, then $\bar{\phi}(x)$ is a local
minimum of the
functional $\tilde{S}[\phi(x)]$.}
\vspace{3mm}

\noindent {\em Proof.}   By construction, the difference
$\tilde{S}[\phi]-S[\phi]= \alpha_1 \mid \Lambda_1 \mid \geq 0$, with equality
possible only if $\Lambda_1[\phi]=0$.  Consider some family of functions
$\phi_\lambda(x), \ 0<\lambda <2$ such that
$\phi_\lambda(x)|_{_{\lambda=1}}=\bar{\phi}(x)$. Then

\begin{equation}
\frac{d}{d\lambda} \tilde{S}[\phi_{\lambda}(x)] \mid_{\lambda=1} =
\alpha_1 \frac{d}{d\lambda} |\Lambda_1[\phi^{(\lambda)}(x)]|
\label{deriv}
\end{equation}
because $\frac{d}{d\lambda} S[\phi_{\lambda}(x)]=0$ at $\lambda=1$.

Consider the following two cases.

1) Suppose $ \frac{d}{d\lambda} \Lambda_1[\phi^{(\lambda)}(x)] \neq 0$.
Since $|\Lambda_1[\phi^{(\lambda)}(x)]| \geq 0$, $|\Lambda_1[\bar{\phi}(x)]|=
\inf~|\Lambda_1[\phi^{(\lambda)}(x)]| =0$.  Therefore, if
$ \frac{d}{d\lambda} |\Lambda_1[\phi^{(\lambda)}(x)]| \neq 0$, then
$\bar{\phi}(x)$ is a local minimum of $\tilde{S}[\phi^{(\lambda)}(x)]$ with
respect to $\lambda$.

2) If, on the other hand, $ \frac{d}{d\lambda}
|\Lambda_1[\phi^{(\lambda)}(x)]|_{\lambda=1} = 0$, then

\begin{eqnarray}
0=\frac{d}{d\lambda}\Lambda_1[\phi^{(\lambda)}(x)]_{\lambda=1} =
\frac{d}{d\lambda} (T[\phi^{(\lambda)}(x)]+2V[\phi^{(\lambda)}(x)])
\mid_{\lambda=1} = \nonumber \\
\frac{d}{d\lambda} (S[\phi^{(\lambda)}(x)]+V[\phi^{(\lambda)}(x)])
\mid_{\lambda=1} =
\frac{d}{d\lambda} V[\phi^{(\lambda)}(x)]_{_{\lambda=1}}
\label{Vconst}
\end{eqnarray}
or, in other words, $\Lambda_1$ can have a zero derivative only in the
direction of the constant $V$. However, it was shown in Ref. \cite{c_comm}
that the bounce $\bar{\phi}(x)$ is the global minimum of the action $S[\phi]$
over the surface of constant $V$.

In other words, the variation $(\frac{d}{d\lambda} \phi^{(\lambda)}(x))
d\lambda$
is tangential to the surface $V=const$, while the only \cite{c_comm}
eigenvector of the
second derivative operator $(\delta^2 S/\delta \phi_i \delta \phi_j)=
\delta_{ij} \Delta+ (\partial^2 U(x)/\partial \phi_i \partial\phi_j)$ which
corresponds to
a negative eigenvalue was shown in Ref. \cite{c_comm} to be orthogonal to
the surface $V=V[\bar{\phi}]=const$ in the functional space.  Therefore,
$\bar{\phi}$ is a minimum of $\tilde{S}$ in this case as well.

We have shown that for an arbitrary variation $\phi^{(\lambda)}(x)$ near
the bounce $\bar{\phi}(x)$, either (i) $\delta S=0, \ \delta
|\Lambda_1|^p>0$, or
(ii) $\delta S=\delta \Lambda_1=0$, but $\delta^2 S/\delta \phi^2$ is
positive definite. In either case, $\bar{\phi}$ is the minimum of
$\tilde{S}$.  This completes the proof.  $\Box$

Therefore, the improved action (\ref{st}), as well as a more general form
(\ref{imprvd}) has the desired solution $\bar{\phi}(x)$ as its minimum.  This
justifies the method described above of searching for the bounce as a
minimum of the improved action rather than a saddle point of the actual
action.

Our method can be easily generalized to the case of finite-temperature
field theory.  In the high-temperature limit, the  transition probability
\cite{linde},

\begin{equation}
\Gamma/{\sf V}= T \left ( \frac{S_3[\tilde{\phi},T]}{hT}\right)^{3/2}
e^{-S_3[\tilde{\phi},T]/\hbar T}
\left |
\frac{\det'[-\partial_\mu^2+U''(\tilde{\phi},T)]}{\det[-\partial_\mu^2+
U''(0,T)]}
\right |^{-1/2}
\times (1+O(\hbar)),
\end{equation}
is determined by the three-dimensional action $S_3(\tilde{\phi},T)$ of the
O(3)-symmetric solution $\tilde{\phi}(x)$ of the equation $\delta S_3=0$.
As before, $\tilde{\phi}(x)$ is a saddle point of $S_3[\phi]$.
The improved action $\tilde{S}_3[\phi]$, for which $\tilde{\phi}(x)$ is a
minimum, can be constructed by analogy with equation (\ref{imprvd}):

\begin{equation}
\tilde{S}_3[\phi]:= S_3[\phi]+ \sum_{n} (\alpha_{1,n} |\Lambda^{(T)}_1|^{p_n}+
\alpha_{2,n} |\Lambda^{(T)}_2|^{p_n}+\alpha_{3, ij , n} |\Lambda^{(T)ij}_3
|^{p_n}); \ \ p_n, \alpha_n>0
\label{imprvd_T}
\end{equation}
where

\begin{eqnarray}
\Lambda^{(T)}_1 & = & T+3V \label{lT1} \\ & & \nonumber \\
\Lambda^{(T)}_2 & = & \sum_{n} (6-n) V_n  \label{lT2} \\ & &
\nonumber \\
\Lambda^{(T)ij}_3 &  = & \int d^3 x \  \det
           \left | \begin{array}{cc}
                    \frac{\partial}{\partial \phi_i} &
\frac{\partial}{\partial \phi_j}\\
                    \phi_i &  \phi_j
                   \end{array}  \right | U(\phi_1, ... ,\phi_n,T)
\mid_{\phi(x)=\bar{\phi}(x)}  \label{lT3}
\end{eqnarray}

The application of our method to the issue of stability of the charge and
color conserving vacuum in the MSSM will be presented in \cite{kls}.

The author would like to thank P. Langacker, A. Linde, G. Segre and P.
Steinhardt for stimulating discussions.  This work was supported by the
U.~S. Department of Energy Contract No. DE-AC02-76-ERO-3071.

\end{document}